\documentstyle[12pt]{article}
\newcommand{\bea}{\begin{equation}}
\newcommand{\eea}{\end{equation}}
\newcommand{\ber}{\begin{eqnarray}}
\newcommand{\eer}{\end{eqnarray}}

\newcommand{\pmx}{\pmatrix}

\begin{document}

\title{Phase instabilities generated by parametric modulation in reaction diffusion systems }
\author
{A.Bhattacharyay\footnote {Email: arijit@fkp.tu-darmstadt.de} \\ Institut f\"ur Festk\"orperphysik,\\ Technische Universit\"at Darmstadt, Hochschulstr.~6,\\ 64289 Darmstadt, Germany\\
and \\ J.K.Bhattacherjee\footnote {Email: tpjkb@mahendra.iacs.res.in}\\
Department of Theoretical Physics,\\
Indian Association for the Cultivation of Science,\\
Jadavpur, Calcutta 700 032, India}

\date{\today}
\maketitle
\begin{abstract}
Effect of external periodic force on an oscillatory order in a reaction diffusion system (Gierer Meinhardt model) has been investigated. The 2:1 resonance situation is found susceptible for the generation of a band of phase instabilities. These phase instabilities, captured on multiple time scales, produces a mismatch between the oscillation frequency of reacting species.  
 \\\\
PACS number(s): 87.10.+e, 47.70.Fw
\end{abstract}

\par
The effect of parametric periodic forcing on oscillatory reaction diffusion systems are being studied with renewed interest to see frequency entrainment and resulting multiphase, steady as well as traveling, orders separated by phase fronts \cite{petrov97,elp98,elp99,linprl00}. Existence of multiphase oscillations are theoretically accounted for by showing the stability of phase separated oscillatory orders in complex Ginzburg Landau equation or in some reaction diffusion models. The mechanism that can possibly cause a slow drift in overall phases of oscillation under periodic forcing, and thus produce stable phase-separated regions, is an important subject for investigation. In view of that, we are going to investigate the effect of periodic forcing in time, on an oscillatory system, on multiple time scales. In almost all reactions diffusion systems, one of the reacting species is dependent on the other for its production and thus does not need be externally supplied. This situation causes a constant phase difference to appear between the homogeneous oscillatory reactants of the system. Here we focus on the possible slow variation of that phase difference as a consequence of varied response to the applied force by the reacting species. The reaction diffusion system we have worked on, is the Gierer-Meinhardt model. The results obtained are the generation of a band of phase instabilities caused by parametric resonance at 2:1 sub-harmonic response to the applied frequency. The instabilities developed at slow time scales and produce a mismatch in frequencies of oscillation of the activator and inhibitor species. The result is interesting because in such a situation when activator and inhibitors are oscillating with different frequencies, many different things can follow. An incommensurency in the activator and inhibitor frequencies of oscillation can possibly generate weak phase chaos. In other case, if one of the two species tries to lock itself in phase with the other, a slow drift in the overall phase of oscillation can presumably be an outcome. In order to get rid of such frequency mismatch at various orders, a finer adjustment of the parameter is needed. Such an adjustment also causes a nontrivial shift in phase boundary obtained from the linear stability analysis of the system. 
\par
 The Gierer-Meinhardt model that we have taken up is  \cite{koc,Gie} \ber\nonumber
\frac{\partial
A}{\partial {t}}&=&D{\bigtriangledown} ^2 A +  \frac{A^2}{B} - A +\sigma \\
\nonumber\frac{\partial B}{\partial {t}}&=&{\bigtriangledown}
^2 B +\mu (A^2 - B) \\
\eer
$D$ is the diffusivity of the activator $A$ and is always less than unity to satisfy the Turing condition. The $\sigma$ is the basic production rate of activator where $\mu$ can be interpreted as the production constant and at the same time removal rate of the inhibitor $B$. 
 The linear stability analysis of this model shows that 
the homogeneous stationary basic fixed point $B=A^2=(1+\sigma)^2$ which becomes unstable \cite{koc}
to a time independent spatially inhomogeneous state when
\ber\nonumber \mu D \le
\left(\sqrt{\frac{2}{1+\sigma}} - 1\right)^2 \\
\eer and to a time periodic spatially homogeneous state if
\ber\nonumber \mu < \frac{1-\sigma}{1+\sigma} \\
\eer For the time periodic state the oscillation frequency is \cite{ari} \ber
\nonumber \omega_0 = \sqrt{\frac{1-\sigma}{1+\sigma}}\\ \eer 

Figure 1. shows the phase boundaries in the removal rate $\mu$  vs. diffusivity $ D$ space. On the left of the broken curve and above the continuous horizontal line the Turing state is stable. In what follows, we will concentrate on the region below the continuous horizontal line in Fig.1 where a Hopf state is available.

\par 
Here we are going to consider the problem which is often treated in hydrodynamic instabilities - the one where the control parameter is given a sinusoidal temporal variation \cite{jkb,ahl,hall,fau}. To do so we first linearize Eq.(1) around the fixed point
$B=A^2=(1+\sigma)^2$ and near the boundary $ \mu =
\frac{1-\sigma}{1+\sigma}$. Thus we have \ber\nonumber L_0\pmx{\delta A
\cr \delta B \cr} = 0 \\
\eer where the operator $L_0$ reads \ber\nonumber L_0 =
\pmx{\frac{\partial}{\partial t}-\frac{1-\sigma}{1+ \sigma} &
\frac{1}{(1+\sigma)^2} \cr -2\mu(1+\sigma) &
\frac{\partial}{\partial t}+\mu \cr} \\
\eer We note that the state obtained for $\mu <
\frac{1-\sigma}{1+\sigma}$ is spatially homogeneous but temporally oscillating. We now endow
$\mu$ with a small amplitude time modulation \ber\nonumber \mu =
\mu_0\left(1+\epsilon Cos(\omega t)\right) \\
\eer and ask the question: what is the critical value of $\mu_0$
for which the spatially homogeneous oscillatory state is formed.
\par The linearized equation now reads \ber\nonumber
L_0\pmx{\delta A \cr \delta B \cr} = \epsilon Cos(\omega t)\pmx{0
& 0 \cr 2\mu_0(1+\sigma) & - \mu_0 \cr}\pmx{\delta A \cr \delta B
\cr}\\
\eer The critical value of $\mu_0$ can be expanded as
\ber\nonumber \mu_0 = \mu_{00} + \epsilon\mu_{01} + \epsilon^2\mu_{02} + .... \\
\eer where $\mu_{00}=\frac{1-\sigma}{1+\sigma}$ and in $L_0$ it is
implied that $\mu = \mu_{00}$. In $O$(1), the eigenvector for the homogeneous
oscillatory state below the afore said Hopf-bifurcation boundary
is 
\ber\nonumber \pmx{\delta A_0 \cr \delta B_0 \cr}=\pmx{1 \cr
\frac{2\mu_{00}(1+\sigma)}{\mu_{00}+i\omega_0} \cr}e^{i\omega_0 t} + c.c.
\\ \eer 
 Thus we see that $\delta A_0$ and $\delta B_0$ has a
constant phase difference since the production of ${ B}$ depends on ${ A}$. At this point we
consider this Phase difference $\phi(\mu)$ has an additive part
which varies on a slower time scale $\epsilon\tau$. So the
structure of $\phi(\mu)$ is taken as \ber\nonumber \phi = \phi_c +
\delta\phi(\tau ) \\\eer where $\phi_c$ is the critical value of
$\phi(\mu)$ and can be easily obtained from Eq.(10). The $\delta\phi$ is expanded in powers of $\epsilon$
as \ber\nonumber \delta\phi = \delta\phi_0 + \epsilon\delta\phi_1
+ \epsilon^2\delta\phi_2 + ....\\\eer Let us expand $\delta A$ and
$\delta B $ as \ber\nonumber \delta A =
\delta a_0 + \epsilon\delta a_1 +\epsilon^2 \delta a_2 + .... \\
\nonumber \delta B = \delta b_0 + \epsilon \delta b_1 +
\epsilon^2\delta b_2 + ....
\\ \eer  and introduce the multiple time scales as
\ber\nonumber t = t_0 + \epsilon\tau \\
\eer where \ber\nonumber \frac{\partial}{\partial t} =
\frac{\partial}{\partial t_0}+ \epsilon\frac{\partial}{\partial
\tau}\\
\eer We now specialize to the case of parametric resonance i.e.
$\omega = 2\omega_0$. The $O(\epsilon)$ equation is \ber\nonumber L_0\pmx{\delta a_1 \cr \delta b_1 \cr} =
\pmx{0 \cr \delta b_+\frac{\partial}{\partial \tau}(\delta\phi_0)
+ \mu_{01}[2(1+\sigma)\delta a_+ - b_+] +\frac{\mu_{00}}{2} [2(1+\sigma)
\delta a_- -\delta b_-] \cr} \\
\eer On the right hand side of above equation, only the secular
terms have been considered. The $\pm$ sign indicates $\pm i\omega$ in the appropriate expressions of amplitudes $\delta a_0$ and $\delta b_0$ of the $O(0)$ solution. Now, the condition of occurrence of
$O(\epsilon)$ solution i.e. the vanishing of secular terms lead
to \ber\nonumber\frac{\partial}{\partial \tau}(\delta\phi_0) =
-\frac{\mu_{01}[2(1+\sigma)\delta a_+ - b_+] +\frac{\mu_{00}}{2}
[2(1+\sigma) \delta a_- -\delta b_-]}{\delta b_+}\\ \eer Coming back to the original time scale and simplifying \ber\nonumber\frac{\partial}{\partial t}(\delta\phi_0) = -\frac{i\epsilon\omega_0}{\mu_{00}}\left(\mu_{01}
- \frac{\mu_{00}(\mu_{00}-1)}{2(\mu_{00}+1)}\right) -\frac{\epsilon\omega_0^2\mu_{00}}{
\mu_{00}^2+\omega_0^2} \\ \eer The condition that there will occur $O(\epsilon)$ solution is the occurrence of a mismatch in frequencies of oscillation of activator and inhibitor by the amount $ \frac{\omega_0}{\mu_00}\left(\mu_1
-\frac{\mu_0(\mu_0-1)}{2(\mu_0+1)}\right)$. One has to pull the Hopf boundary up by the amount \ber\nonumber\mu_{01}
=\frac{\mu_{00}(\mu_{00}-1)}{2(\mu_{00}+1)}\\\eer to do away with such frequency mismatch between activator and inhibitor or in other words to get rid of weak disturbances. 
\par The solution of $O(\epsilon)$ equation for $( \omega =
2\omega_0)$ will now
definitely include $e^{\pm i3\omega_0}$ term. This term will make
secular terms appear in the next higher order. The $O(\epsilon)$
solution in $e^{\pm i3\omega_0}$ will come as
\ber\nonumber \pmx{\delta {\bar{a_1}} \cr \delta {\bar{b_1}}\cr} =
\frac{\mu_{00}}{2\Delta_3\omega_0}[2(1+\sigma)\delta a_+ -\delta
b_+]\pmx{\frac{-1}{(1+\sigma)^2} \cr i3\omega_0 - \mu_{00}
\cr}e^{i3\omega_0 t} + c.c. \\ \eer The $\Delta_{3\omega_0}$ being
the determinant of $L_0$ at frequency $3\omega_0$. Let us look at the
$O(\epsilon^2)$ equation with only the secular part (i.e. the part
of frequency $\omega_0$) on the right hand side. \ber\nonumber
L_0\pmx{\delta a_2 \cr \delta b_2 \cr} = \pmx{0 \cr \delta
b_+\frac{\partial}{\partial \tau}(\delta\phi_1) +
\mu_{02}[2(1+\sigma)\delta a_+ - b_+] +\frac{\mu_{00}}{2} [2(1+\sigma)
\delta {\bar{a_1}} -\delta{\bar{b_1}}] \cr} \\
\eer Again the removal of secular terms results in  \ber\nonumber
\frac{\partial \phi_1}{\partial t} &=&
-\epsilon\frac{\omega_0}{\mu_{00}}
\left[i\left(\mu_{02}+(\frac{\mu_{00}}{2})^2\frac{1}{\Delta_{3\omega_0}}\right)-(\frac{\mu_{00}}{2})^2\frac{3}{{\Delta_{3\omega_0}}}\right] \\ \eer So the condition
for existence of $O(\epsilon^2)$ solution is a further mismatch of frequency of oscillation of the reacting species. A fine tuning of $\mu $ by \ber\nonumber \mu_{02} =
(\frac{\mu_{00}}{2})^2\frac{1}{\mu_{00}^2+9\omega_0^2
-\frac{2(1+\sigma)}{(1+\sigma)^2}}
\\\eer can help getting rid of this frequency mismatch at this order too. In the above expression we have put
\ber\nonumber {\Delta_{3\omega_0}} = -({\mu_{00}^2+9\omega_0^2
-\frac{2(1+\sigma)}{(1+\sigma)^2}})\\\eer

 Thus we
see that persistence to retain the frequency $\omega_0 $ for the
$O(\epsilon^2)$ solution causes a further shift in the Hopf-bifurcation boundary. \par
The $O(\epsilon^2)$ solution will also have a $e^{\pm i3\omega_0
t}$ part resulting from $\mu_1[2(1+\sigma) \delta \bar{a_1}
-\delta\bar{b_1}]$ which will cause secular term to appear in the
immediate higher order. The above mentioned part of the solution
in $O(\epsilon^2)$ is of the form \ber\nonumber \pmx{\delta
{\bar{a_2}} \cr \delta {\bar{b_2}}\cr} =
\frac{\mu_{01}}{2\Delta_3\omega_0}[2(1+\sigma)\delta\bar{a_1} -\delta
\bar{b_1}]\pmx{\frac{-1}{(1+\sigma)^2} \cr i3\omega_0 - \mu_{00}
\cr}e^{i3\omega_0 t} + c.c. \\\eer This can be written in a simplified form as
\ber\nonumber
 \pmx{\delta
{\bar{a_2}} \cr \delta {\bar{b_2}}\cr} = -\frac{\mu_{01}}{2\Delta_3\omega_0}(i3\omega_0 +1)\pmx{\delta
{\bar{a_1}} \cr \delta {\bar{b_1}}\cr}\\\eer
Again the $O(\epsilon^3)$ solution will contain a
$e^\pm{i3\omega_0 t}$ part coming from the term
$\mu_1[2(1+\sigma)\delta\bar{a_2}-\delta\bar{b_2}]$ resulting in
\ber\nonumber
 \pmx{\delta
{\bar{a_3}} \cr \delta {\bar{b_3}}\cr} = \left[-\frac{\mu_{01}}{2\Delta_3\omega_0}(i3\omega_0 +1)\right]^2\pmx{\delta
{\bar{a_1}} \cr \delta {\bar{b_1}}\cr}\\\eer and so on. In this
way at $O(\epsilon)^n$  \ber\nonumber
 \pmx{\delta
{\bar{a_n}} \cr \delta {\bar{b_n}}\cr} = \left[-\frac{\mu_{01}}{2\Delta_3\omega_0}(i3\omega_0 +1)\right]^{(n-1)}\pmx{\delta
{\bar{a_1}} \cr \delta {\bar{b_1}}\cr}\\\eer 
Thus we get a nontrivial flow of secular term generators at all orders producing phase instabilities. Such a band of phase instability is generated as a result of sub-harmonic response to the external force. These instabilities would possibly show up in weak phase turbulence under forcing.

\par
In the conclusion we would like to mention that, a multiple scale perturbation analysis of forced Hopf state reveals the presence of instabilities which are responsible for a slow frequency drift between the two reacting species. We would like to focus on the point of instantaneous varied response of activator and inhibitors is the basic cause of generation of such instabilities. This type of differential response to an applied force can always occur when one of the reactants depends on the other for its production and thus allowing for a delay. A persistence of the reacting species for oscillating in unison can cause in a slow overall phase drift of the system. Such oscillatory regions separated by continuous distribution of relative phases has been experimentally observed by Lin et al. in the forced Belousov-Zhabotinsky system at low forcing amplitude \cite{linprl00}. At high enough forcing amplitude they have got well defined $\pi $ phase separated 2:1 resonant pattern. It is important to note that a gradual increase in the parameter $\mu $, in our analysis, to get rid of spurious instabilities is also the same as enhancing the forcing amplitude. Consideration of large scale spatial phase variations under resonant forcing can easily be shown to result in an inhomogeneous diffusion equation in $\phi $ with a complex inhomogeneity $ C $ originating as a result of external forcing. This type of a situation can result in spatial instabilities in one of the reacting species. Thus other spatio-temporal instabilities at larger scales can also be shown to exist, which influences the situation under forcing.

\newpage
{\bf Figure caption} \\
Figure 1 shows the phase diagram of the one dimentional Gierer-Meinhardt model on removal rate $\mu$ vs. diffusivity $D$ space. The continuous line is a Hopf bifurcation boundary whereas the broken line separates steady Turing state (on the left) from the basic homogeneous steady state.
\end{document}